\documentclass[10pt,showkeys,showpacs,amsmath,amssymb,prl]{revtex4}
\usepackage[english]{babel}
\usepackage[dvips]{graphicx}
\usepackage{amsmath,euscript,amssymb}
\newcommand\ba{\begin{eqnarray}}
\newcommand\ea{\end{eqnarray}}
\newcommand\be{\begin{equation}}
\newcommand\ee{\end{equation}}
\newcommand\nn{\nonumber}

\begin{document}
\title{About the role of vector mesons on
the $\eta \to \pi^0 \gamma \gamma$ decay width   in
  meson -- baryon chiral model
 }

\author{E. A.  Kuraev}%
\email{kuraev@theor.jinr.ru}
\author{V. N.  Pervushin}%
\email{pervushin@theor.jinr.ru}
\author{M. K. Volkov}
\email{volkov@theor.jinr.ru}
\affiliation{%
Bogoliubov Laboratory of Theoretical Physics, Joint Institute for Nuclear Research,
141980 Dubna, Russia }

\date{\today}
\begin{abstract}
It is  shown  in the work of one of the authors in 1979 (MKV) 
that  the contribution to the amplitude of this decay from
diagrams with one baryon loop  is equal to zero and contributions
from  diagrams with meson loops  appear very small. However, pole
diagrams with intermediate vector mesons were not  considered
there. Here it is shown that  contributions of these pole diagrams
dominate.
  The meson-baryon chiral model used here is compared with
known quark chiral models. The obtained results are in
satisfactory agreement with recent experimental data.


\end{abstract}
\maketitle

The investigations of the process $\eta \to \pi^0 \gamma \gamma$
have a long history \cite{1}-\cite{8} (see in detail  \cite{1}).
The first theoretical estimates of a depth of this process arose
in the sixties of the last century \cite{1}. The first
experimental data were based on a few statistics and they were not
precise. So  these results gave arguments  to predict a large
value of the branching ratio of the process $26$ eV. Critical
arguments of this prediction can be  found in \cite{2}. A real
breakthrough in the investigation of this process happened in the
GAMS experiment in 1981 at Protvino \cite{3} where the value
$\Gamma_{\eta \to \pi \gamma \gamma}=0.84 \pm 0.18$ eV was
obtained. Notice that this result is consistent with those
obtained in the Nambu -- Jona-Lasinio model  \cite{4,5}.

 At the last time,  attention to this process
 revived from both a theoretical and experimental points
 of view (see \cite{7} and cites in it).  It was promoted first
 of all by careful measurement in the CERN laboratory  of
 the $\eta \to 3\pi$ and $\eta \to
 \pi_0\gamma\gamma$ decay probabilities,
 where about one million events was already  investigated.
 As a result, the estimation of the width  decreased in
 comparison with Prokoshkin's estimation
  \cite{3} almost twice. During the same time theoretical
works were published in which an attempt to  explain these results
was undertaken.
   In these works \cite{6} vector dominance models,
various chiral quark models or a model with meson loops were used
in the framework of Chiral Perturbation Theory (ChPT) with
decomposition into momenta down to
the eighth order  $p^8$.

Let us pay attention to that in quark chiral models there is a
number of basic lacks. Namely, by consideration of quark loops the
quark confinement principle   was not  provided. Ultra-violet
divergence was eliminated by introduction of a limiting momentum.
In using linear sigma models there is a problem of  explanation of
the big mass $a_0(980)$ meson in the framework of both the chiral
quark model and Nambu -- Jona-Lasinio (NJL) one \cite{5}.

\begin{figure}
\includegraphics[scale=1.2]{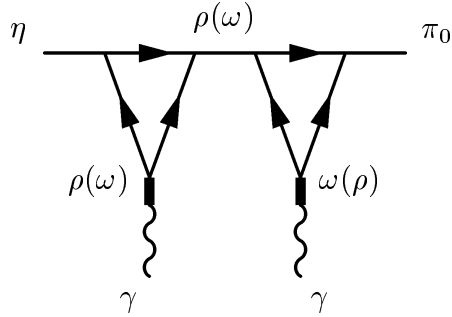}
\caption{Feynman diagrams with intermediate vector mesons.}
\label{fig:1}
\end{figure}

By virtue of the above -- said of  certain interest is the
description of this process by using  the chiral meson-baryon
Lagrangian by analogy with the works of 1978-1979 \cite{2,9}. In
 \cite {2}, the contribution of intermediate vector $\omega$ and
$\rho$ - mesons (see fig. \ref{fig:1}) however was not  considered.

The absence of confinement problems is thought to be the advantage
of the chiral meson-baryon model. An important point is the
absence of ultra-violet divergences, which is a consequence of
gauge invariance  of the model. Further, in view of large baryon
masses in calculating  amplitudes corresponding to loop diagrams,
it is possible to use only  the lowest terms of decomposition into
external momenta. We show in this work the approximate validity of
the principle of  quark-hadron duality (see also paper of Volkov
and Osipov in ref. [6]). In the description in
terms of quark  models independence of triangular amplitudes of
the constituent quark mass is used and the factor three-number
colors is introduced. In our approach it is necessary to use the
Goldberger-Treimann relation for a constant of meson-baryon
coupling $g=g_A M_B/F_\pi, F_\pi=93 MeV$, which includes the
constant of renormalization of a  strong vertex top in the neutron
$\beta$ decay  $g_A=1,25$,  consider contributions of charged
baryons from an octet with the corresponding factors following
from $SU (3)$ symmetry and consider their various masses.
Dependence on the baryon masses $M_B $ in amplitudes corresponding
to diagrams of abnormal type
 disappears  as  in the case of quarks. As a result, we
obtain the factor $3.3 $ instead of the color factor 3 quarks
accepted in the model for  the triangular diagram $ \pi_0 \rho
\omega $.

This "almost full" coincidence can be  considered as the proof of
the validity of the principle of  the quark-hadron duality. We
show that within the framework of the realistic chiral-symmetric
baryon-meson model, where physical observable objects are only
used, it is possible to get a satisfactory description of
radiation decays of vector mesons  and a rare process $ \eta \to
\pi_o \gamma \gamma $ in accord with recent experimental data. It
is the motivation of our work.

In the beginning, using  part of the baryon-meson interaction
Lagrangian, we calculate correcting factors for the anomalous
triangular diagrams which will play an important role in the
following calculation. Using these factors and results obtained
within the  quark models, we describe the widths of radiation $
\rho (\omega) \to \pi_0 (\eta) \gamma $ vector meson decays. Also,
we compare these widths with  modern experimental data. Further we
will consider double radiation $\eta$ decay  $\eta \to \pi_0
\gamma \gamma $. In the conclusion, discussion of the obtained
results is given.

Part of the Lagrangian describing interaction of an octet of
baryons with neutral pion and $\eta$-meson  looks like [2,9]
 \ba
L_{P\bar{N}N}=ig\pi_0\left[\bar{p}\gamma_5
p+\frac{1}{3}\bar{\Xi}^-\gamma_5 \Xi^-+\frac{2}{3}
(\bar{\Sigma}^+\gamma_5 \Sigma^+-\bar{\Sigma}^-\gamma_5 \Sigma^-)\right]+ \nn \\
i\frac{g}{\sqrt{3}}\eta_8\left[\frac{1}{3}\bar{p}\gamma_5 p-
\frac{5}{3}\bar{\Xi}^-\gamma_5 \Xi^-+\frac{4}{3}
(\bar{\Sigma}^+\gamma_5 \Sigma^++\bar{\Sigma}^-\gamma_5 \Sigma^-)\right]+ \nn \\
ig\eta_0\left[\bar{p}\gamma_5 p+\bar{\Xi}^-\gamma_5
\Xi^-+\bar{\Sigma}^+\gamma_5 \Sigma^+ +\bar{\Sigma}^-\gamma_5
\Sigma^-\right], \ea where $\pi_0$ is the pion field,
$\eta_0,\eta_8$ are the singlet and octet components of  $\eta$-
meson, $p,\Xi^-,\Sigma^\pm$ are the baryon octet.  For
completeness let us also give Lagrangians of interactions of
vector mesons with baryons \ba
L_{V\bar{N}N}=i\frac{g_\rho}{2}[\rho_\mu\left(\bar{p}\gamma_\mu
p-\bar{\Xi}^-\gamma_\mu \Xi^-+ \bar{\Sigma}^+\gamma_\mu
\Sigma^+-\bar{\Sigma}^-\gamma_\mu \Sigma^-
\right)+ \nn \\
+\omega_\mu\left(\bar{p}\gamma_\mu p+\bar{\Xi}^-\gamma_\mu \Xi^-+
\bar{\Sigma}^+\gamma_\mu \Sigma^++\bar{\Sigma}^-\gamma_\mu
\Sigma^-\right)], \ea and ones describing  photon-vector meson
transitions  (vector dominance model) \ba
L_{VA}=\frac{e}{g_\rho}A_\mu\left[m_\rho^2\rho_\mu+\frac{1}{3}m_\omega^2\omega_\mu+
\frac{\sqrt{2}}{3}m_\phi^2\phi_\mu\right], \ea where $A_\mu$ is a
photon field, and $\rho_\mu,\omega_\mu, \phi_\mu$ are the nonet
vector fields.

Let us calculate correcting factors which are to be considered in
transition from quark models to the model with baryons. For the
triangular diagram with external $ \pi \rho \omega $, using the
resulted formulas, we have  \ba
K_\pi=\frac{1}{3}g_A\left[1+\frac{1}{3}+\frac{4}{3}\right]=1.1.
\ea

For the vertex functions $\eta\omega\omega,\eta\rho\rho$ we get
\ba
K_{\eta_8}=\frac{1}{3\sqrt{3}}g_A\left[1-\frac{5}{3}+\frac{8}{3}\right]=0.326 ;\nn \\
K_{\eta_0}=\frac{1}{3}g_A[1+1+2]=1.68. \ea

As a result, for real  $\eta$-meson,  taking into account  the
singlet-octet mixing
$$\eta=\eta_0\cos(\theta_0-\theta)+\eta_8\sin(\theta_0-\theta),$$
where  an ideal mixing angle is $\cos\theta_0=\sqrt{2/3}$, and
choosing a deviation from an ideal angle $\theta=-18^0$, we obtain
\ba K_\eta=1.27. \ea

Expressions for the  decay widths of  vector mesons, given in [5],
in view of the correcting factors look like
 \ba
\Gamma^{\rho\to \pi_0\gamma}=K_\pi^2\frac{2\pi\alpha}{3}
\frac{g_\rho^2}{4\pi}\frac{1}{(16\pi^2F_\pi)^2}
\left(\frac{M_\rho^2-M_\pi^2}{M_\rho}\right)^3, \nn \\
\Gamma^{\omega\to \pi_0\gamma}=9\Gamma^{\rho\to \pi_0\gamma}; \ea
\ba
\Gamma^{\rho\to \eta\gamma}=K_\eta^2\pi\alpha
\frac{g_\rho^2}{4\pi}\frac{1}{(16\pi^2F_\pi)^2}
\left(\frac{M_\rho^2-M_\eta^2}{M_\rho}\right)^3, \nn \\
\Gamma^{\omega\to \eta\gamma}=\frac{1}{9}\Gamma^{\rho\to
\eta\gamma}. \ea
 Substituting the above  factors  and also
$F_\pi=93 MeV,g_\rho=5.94$ we have \ba
\Gamma_{th}^{\rho\to \pi_0\gamma}=93.5 KeV; \nn \\
\Gamma^{\omega\to \pi_0\gamma}_{th}=930 KeV; \nn \\
\Gamma^{\rho\to \eta\gamma}_{th}= 42.8 KeV;\nn \\
\Gamma^{\omega\to \eta\gamma}_{th}= 4.96 KeV. \ea
 For comparison we write out the results of experiments  (PDG-2006): \ba
\Gamma^{\rho\to \pi_0\gamma}_{exp}=(89.4\pm 12)KeV; \nn \\
\Gamma^{\omega\to \pi_0\gamma}_{exp}=(756,5\pm 21)KeV; \nn \\
\Gamma^{\rho\to \eta\gamma}_{exp}= (44.7\pm4.5)KeV; \nn \\
\Gamma^{\omega\to \eta\gamma}_{exp}=(4.17\pm0.5)KeV. \ea Comparing
these results, we can see that they are in  satisfactory agreement
with each other. We  now  consider the process
$\eta\to\pi_0\gamma\gamma$.

As shown in work  [2] the contributions of amplitudes to Feynman
diagrams corresponding to one-baryon loop with three and four
vertices within the limits of chiral baryon  models precisely
compensate each other,  the contribution of diagrams with mesons
in the closed loops is the value of  an order of
$\Gamma_{(\pi)}\approx 0.01eV$  and, thus, is very small.

Hence it follows that in  describing the decay
$\eta\to\pi_0\gamma\gamma$ the dominating contribution comes from
the diagrams with intermediate $\rho,\omega$ mesons.

Calculation of the corresponding contribution is close to a
similar calculation within the framework of quark models [5]. In
[5] the factor $\cos\theta-\sqrt{2}\sin\theta\approx 1.39$ was
 used for $\theta=-18^0$ which describes the singlet-octet mixing
for $\eta$-meson. It is necessary to replace this factor by
$K_\eta=1.27$, obtained above. Using the results of numerical
calculations [5] we get for the width of double radiation decay of
this meson
\ba
\Gamma^{\eta\to\pi_0\gamma\gamma}_{th}=0.326~\mbox{eV}.
\ea
This result is in satisfactory agreement with the result of
experimental work [8]
 \ba
\Gamma^{\eta\to\pi_0\gamma\gamma}_{exp}=(0.45\pm 0.12)~\mbox{eV}.
\ea

We are grateful to S.B. Gerasimov and A.E. Dorokhov for useful
discussions, V. Bytev and N. Kornakov for help. One of us (MKV) acknowledges
the support of the Russian Foundation for Basic Research
05-02-16699, EAK is grateful to the INTAS-05-1000008-8328.

\end{document}